\documentclass[12pt,a4paper]{article}

\usepackage[usenames,dvips]{color}

\usepackage[latin2]{inputenc}
\usepackage{epic}
\usepackage{latexsym}
\usepackage{epsf}
\usepackage{epsfig}
\usepackage{amssymb,amsmath}

\def\be{\begin{equation}}
\def\ee{\end{equation}}
\def\bea{\begin{eqnarray}}
\def\eea{\end{eqnarray}}

\def\nn{\nonumber}
\def\al{&&\!\!\!\!\!\!\!\!}
\def\d{{\rm d}}
\def\sgn{{\rm sgn}}

\def\ZZ{\mathbb Z}
\def\cO{{\cal O}}
\def\ddn{\frac{\partial}{\partial n}}

\begin{document}
\thispagestyle{empty}
\begin{flushright}
IFT--08--01\\                  
\end{flushright}

\vspace*{1cm}

\begin{center}
{\Large\bf k-stabilization in brane models}
\vspace*{5mm}
\end{center}
\vspace*{5mm} \noindent
\vskip 0.5cm
\centerline{\bf 
M. Olechowski
}
\vskip 5mm
\centerline{\em Institute of Theoretical Physics, 
University of Warsaw}
\centerline{\em ul.\ Ho\.za 69, PL--00--681 Warsaw, Poland}

\vskip 1cm

\centerline{\bf Abstract}
\vskip 3mm
Stabilization of inter--brane distance is analyzed in 
5--dimensional models with higher--order scalar kinetic terms.
Equations of motion and boundary conditions for background 
and for scalar perturbations are presented. 
Conditions sufficient and (with one exception) necessary for 
stability are derived and discussed. It is shown that it is possible
to construct stable brane configurations even without  
scalar potentials and cosmological constants.
As a byproduct we identify a large class of non--standard 
boundary conditions for which the Sturm--Liouville operator 
is hermitian.

\newpage
\section{Introduction}

Higher dimensional brane models belong to the most interesting 
recent developments in the theory of fundamental interactions.
Many models have been proposed in which the space--time consists 
of a 5--dimensional (5D) bulk ending at two 4--dimensional (4D) 
branes. Usually this space--time has the structure of a warped 
product of a maximally symmetric 4D space--time and the one 
dimensional orbifold $S^1/\ZZ_2$ with the branes located at the 
$\ZZ_2$ fixed points. The Standard Model fields may propagate only 
on one of the branes called the visible one. Some other fields may 
live on the second, hidden, brane. Of course, the gravity fields 
can propagate in the whole 5D space--time.

Phenomenological features of such models depend on  
fields and interactions other than that of the Standard Model, 
on the warping, and on the distance between the branes. 
This distance must be fixed in a stable way. Such stabilization 
can not be achieved with only gravity propagating in the bulk. 
A simple mechanism of fixing the inter--brane distance was
proposed by Goldberger and Wise \cite{GoWi}. 
The idea is to add a 5D scalar field with some 
bulk and brane potentials. If the background value of that field 
is not constant in the bulk, then the boundary conditions 
(or in another words: equations of motion at the branes) 
can be fulfilled only if the branes are located at appropriate 
points in the 5th dimension.

It is not enough to have a background solution with some fixed 
brane positions. It is necessary also that such a configuration 
is stable against all possible small perturbations. 
From the 4D point of view, the perturbations can be 
describe in terms of Kaluza--Klein (KK) towers of states. 
The lightest scalar KK state is usually called the radion
\cite{ChGrRu}. 
Tachyonic character of the radion indicates instability 
of a given background. The problem of the radion mass, or 
of the stability of the inter--brane distance, was investigated 
by many authors \cite{DWFrGuKa,TaMo,CsGrKr,MuKo,LeSo}. 
Its relation to inflation was discussed in \cite{FrKo}. 
Quite general criteria for the stability were found in \cite{LeSo}.
Generalization of such criteria for models with 
the Gauss--Bonnet interactions was presented in \cite{KoOlSc}.

In the present paper we will do the stability analysis for brane 
models with non--standard kinetic terms for the scalar field. 
Such non--standard kinetic terms appear for example in 
string theory due to the $\alpha'$-- and the loop--corrections.
Very interesting models with generalized scalar kinetic terms 
were investigated in the cosmological context. Kinetically driven 
inflation, called the k--inflation, was introduced in 
\cite{ArDaMu,GaMu}. 
Models of k--essence were proposed as another approach to 
the cosmological constant problem \cite{ChOkYa,ArMuSt}. 
Causality in the context of generalized kinetic terms was 
discussed by many authors (see e.g.\ \cite{BaMuVi} and 
references therein).

There is a simple reason to expect that models with non--standard 
scalar kinetic terms may be interesting for the inter--brane 
distance stabilization. Their Lagrangians contain terms with 
more complicated, than just quadratic, 
dependence on the scalar derivatives.
The scalar derivative with respect to the 5th coordinate is 
crucial for the stabilization mechanisms similar to that 
of Goldberger and Wise. This is analogous to the situations 
in cosmological models where the time derivative of the scalar 
field is crucial. The problem of radion stabilization in 
models with non--standard kinetic terms was addressed in 
\cite{DMSS} 
but unfortunately the authors used a method which in general 
is not correct and obtained incorrect 
results\footnote
{
The authors of \cite{DMSS} integrated a Lagrangian with 
fields replaced by their background values. They called the result 
``the effective potential'' and looked for the minima of such an 
object. Of course, in general the potential integrated in a given 
background is not equal to the correct effective potential. 
It happens to be equal in some simple cases, but this must 
be checked case by case by other methods, so the method 
of integrating the potential is practically useless. 
The authors of \cite{DMSS} claim e.g.\ that all models with 
the standard kinetic terms are unstable, what is in clear conflict 
with the results of many previous analyses 
\cite{DWFrGuKa,CsGrKr,LeSo,FrKo,KoOlSc,KoMaPe}.
}.

In addition to the bulk (non-standard) kinetic terms
we will consider also analogous brane-localized ones. 
Many 5D models with (standard) brane kinetic terms 
for different bulk fields were proposed. Such localized 
kinetic terms were investigated for: pure gravity
\cite{DvGaPo,CDLPQW,DvGaKoNi,BCLPS},
gauge fields 
\cite{CaTaWa,Ky,DaHeRi,AgPeSa,AgPeSa2,MuNiPiRu},
fermions \cite{Ky,AgPeSa,AgPeSa2}
and scalars \cite{AgPeSa,CsHuMe}

In section \ref{model} we define our model and derive 
background equations of motion and boundary conditions.
Analogous equations for the scalar perturbations are
presented in section \ref{perturbations}. 
In subsection \ref{self-adjointEP} we show that 
the spectrum of those perturbations is real.  
We identify a large class 
of boundary conditions for which the Sturm-Liouville 
eigenvalue problem is 
self-adjoint.
The stability 
conditions are obtained in section \ref{stability}. They 
are discussed and compared to that in models with the standard 
kinetic terms in section \ref{discussion}. Finally,
section \ref{conclusions} contains our conclusions.

\section{Model and background}
\label{model}

We consider 5D models compactified on the $S^1/\ZZ_2$ orbifold 
with the standard gravitational interactions but with 
non--standard kinetic terms for a scalar field $\Phi$. 
Two 4D branes are localized at the $\ZZ_2$ fixed points $y=y_i$.
The action takes the form
\bea
S=
\int \d^4x\,\d y\,\sqrt{-G}
\al
\left\{
\frac{1}{2\kappa^2}R-P(\Phi,X)-V(\Phi)
\right.
\nn\\
\al
\left.\,\,
-\sum_{i=1}^2 \tilde\delta(y-y_i)
\big[Q^{(i)}(\Phi,X)+U^{(i)}(\Phi)\big]
\right\}
\,,
\label{action}
\eea
where 
\be
X=\frac12 \left(\nabla\Phi\right)^2
\,,
\ee
and $\tilde\delta$ is the normalized Dirac delta satisfying
$\int\d y\,\sqrt{-G}\,\tilde\delta(y-y_i)=\sqrt{-G^{(i)}}$ 
with $G^{(i)}$ being the determinant of the metric induced 
on the brane localized at $y=y_i$ (we chose $y_1<y_2$).
The bulk kinetic term is given by some function $P(\Phi,X)$
depending on the derivatives of $\Phi$ through the combination 
$X$ and on the scalar field itself. We choose $P(\Phi,X)$ in
such a way that it vanishes for $X=0$. This way $V(\Phi)$ 
describes the whole scalar contribution to the action for
constant $\Phi$. In addition to the bulk interactions,  
we consider brane localized 
contributions to the scalar kinetic term and to the potential: 
$Q^{(i)}(\Phi,X)$ and $U^{(i)}(\Phi)$, 
respectively\footnote{
There are two kinds of brane kinetic terms
considered in the literature. 
Some authors assume that such terms 
involve derivatives with respect to all 5 coordinates 
(e.g.\ \cite{AgPeSa}-\cite{MuNiPiRu}) 
while other assume that the derivative in the orbifold 
direction is not present  
(e.g.\ \cite{CDLPQW}-\cite{DaHeRi}). 
We apply the former approach which seems to be natural 
when treating thin branes as limits of thick ones. 
Generalization of our results to the case of brane kinetic 
terms $Q$ which do not depend on $\partial\Phi/\partial y$ 
is quite straightforward.
}
.

The terms in the action (\ref{action}) containing the brane 
localized kinetic functions $Q^{(i)}(\Phi,X)$ must 
be treated with special care.
Let us discuss now in some detail the meaning of 
an integral containing a product of $Q^{(i)}(\Phi,X)$ and
the Dirac delta. Writing explicitly the arguments in one 
of such expressions we get 
\be
\int\d^4 x
\int\d y\, 
\sqrt{G(x,y)}\,
\tilde\delta(y-y_i)\,Q^{(i)}\!\left(\Phi(x,y),
\frac12\left(G^{55}(x,y){\Phi'}^2(x,y)
+\ldots\right)\right)
\label{reg}
,
\ee
where prime denotes differentiation with respect to the 
orbifold coordinate $y$ and, 
in the second argument of $Q^{(i)}$, the ellipsis stand for 
terms in $X$ with derivatives of $\Phi$ in directions
other than $y$.
In brane models the derivatives with respect to the orbifold 
coordinate(s) are usually discontinuous at the brane positions.
The derivative $\Phi'(x,y)$, being a $\ZZ_2$ odd function, 
is exactly zero at $y=y_i$. 
On the other hand, due to the brane sources, the limits 
$\lim_{y\to y_i^\pm}\Phi'(x,y)$ can be different from zero. 
The square of the scalar field derivative, ${\Phi'}^2(x,y)$, 
is even under the $\ZZ_2$ symmetry, and can be written as 
a product of $\sgn^2(y-y_i)$ and a smooth function.
Usually ${\Phi'}^2(x,y)$ is discontinuous at $y=y_i$ and, 
strictly speaking, its integral with the Dirac delta localized 
at $y_i$ is not well defined. All expressions of this kind 
must be regularized. Physically,
such regularization corresponds to using a thick brane and 
taking the limit of its thickness decreasing to zero. 
Technically, one replaces 
$\delta(y-y_i)$ and $\sgn(y-y_i)$ with some smooth functions
$\delta_\varepsilon(y-y_i)$ and $\sgn_\varepsilon(y-y_i)$ satisfying 
the relation $\sgn'_\varepsilon(y-y_i)=2\delta_\varepsilon(y-y_i)$ and 
approaching the Dirac delta and the signum function, respectively, 
when $\varepsilon\to0$.
We calculate the integrals like (\ref{reg}) for regularized expressions 
and at the end remove the regulator taking the limit $\varepsilon\to0$.
Thus, we obtain for example
\be
\int\d^4x \int_{y_a}^{y_b}\d y\, \sqrt{G}\,
\tilde\delta(y-y_i){\Phi'}^{2n}(x,y)
=
\int\d^4x\,\sqrt{G^{(i)}}
\lim_{y\to y_i}
\frac{{\Phi'}^{2n}(x,y)}{2n+1}
\,,
\label{regularization}
\ee
if $y_a<y_i<y_b$. 
It is not necessary to specify the direction of the limit
in (\ref{regularization}) because 
${\Phi'}^{2n}(x,y)$
is an even function of $(y-y_i)$ for any integer $n$.
However, the limit itself is necessary because usually 
${\Phi'}^{2n}(x,y)$
is discontinuous at $y=y_i$.
In the down--stairs approach, one of the limits 
of integration is equal to $y_i$ and the r.h.s.\ of the above 
equation must be multiplied by $1/2$.

In this work we are interested in warped background solutions 
with the flat 4D foliation described by the ansatz
\bea
&\d s^2 = a^2(y)\left
(\eta_{\mu\nu}\d x^\mu\d x^\nu + \d y^2\right),&
\label{ansatz1}
\\[4pt]
&\Phi=\phi(y).&
\label{ansatz2}
\eea
The bulk equations of motion for the system described by 
action (\ref{action}) and satisfying ansatz 
(\ref{ansatz1}--\ref{ansatz2}) 
are given by (we use units $\kappa=1$)
\bea
\label{bulk_bg1}
\left(P_X\phi'\right)'+3\frac{a'}{a}P_X\phi'
-a^2\left(V_\Phi+P_\Phi\right)=0\,,
\\[4pt]
\label{bulk_bg2}
\frac{a''}{a}-2\left(\frac{a'}{a}\right)^2
+\frac{1}{3}P_X{\phi'}^2=0\,,
\\[4pt]
\label{bulk_bg3} 
6\left(\frac{a'}{a}\right)^2
-P_X{\phi'}^2+a^2\left(V+P\right)=0\,, 
\eea
where the subscripts $X$ and $\Phi$
denote derivatives with respect to the arguments  $X$  and $\Phi$, 
respectively\footnote
{
It is straightforward to generalize the equations of motion 
to the case of any non--flat maximally symmetric 4D foliation 
of the 5D background. For example, for the 4D dS space--time
characterized by the Hubble constant $H$, the left hand
sides of equations (\ref{bulk_bg2}) and (\ref{bulk_bg3}) should
be modified by adding $H^2$ and $-6H^2$, respectively.
} 
.

The boundary conditions for the background can be obtained  
from the full equations of motion resulting from (\ref{action})  
with the brane terms taken into account. Integrating such  
equations over an infinitesimal intervals containing the brane 
positions $y_i$ one gets 
\bea
\lim_{y\to y_1^+(y_2^-)}  
a'
=\al 
\left.
\mp\frac{a^2}{6} \left(U^{(i)} + \int\d y\,\delta(y-y_i)Q^{(i)}\right) 
\right|_{y=y_1(y_2)}
\,, 
\label{brane_a}
\\[4pt] 
\lim_{y\to y_1^+(y_2^-)} 
\left(P_X\,\phi'\right) 
=\al
\left. 
\pm \frac{a}{2}
\left(U^{(i)}_\Phi + \int\d y\,\delta(y-y_i)Q^{(i)}_\Phi\right) 
\right|_{y=y_1(y_2)}
\,, 
\label{brane_phi}
\eea 
where for $y_1<y_2$ the upper (lower) signs are to be taken 
for $y=y_1$ ($y=y_2$). From now on we use the Dirac delta  
distribution with the usual normalization $\int\d y\delta(y)=1$ 
(in the upstairs approach).

The above background bulk equations of motion 
(\ref{bulk_bg1}-\ref{bulk_bg3}) and boundary conditions 
(\ref{brane_a}-\ref{brane_phi}) reduce to the know results 
for the standard kinetic terms after substituting 
$P=X$ and $Q^{(i)}=0$.

\section{Scalar perturbations}
\label{perturbations}

Solving the bulk equations of motion (\ref{bulk_bg1}-\ref{bulk_bg3}) 
and the boundary conditions (\ref{brane_a}-\ref{brane_phi}) one can 
find possible background configurations characterized by the warp
factor $a(y)$ and the scalar field $\phi(y)$. Not all such 
background configurations are stable. To check the stability 
one has to consider all possible small perturbations around a given 
background. Instabilities occur if any of the perturbations 
has a tachyonic character. In this paper we concentrate on the
scalar perturbations. From the 4D point of view 
they form an infinite Kaluza--Klein tower of scalars. 
The state with the lowest (4D) mass squared is called the radion.
The positivity of its mass squared is a necessary condition 
for the stabilization of the inter--brane distance.

To find the radion mass we have to investigate the equations of 
motion for the scalar perturbations around the background.
Using the generalized longitudinal gauge, the scalar perturbations
can be written in the following way
\bea
&\d s^2 =
a^2\left[\left(1+2F_1\right)
\left(\eta_{\mu\nu}\d x^\mu\d x^\nu\right)
+\left(1+2F_2\right)\d y^2\right],
&
\label{ansatz_F1F2}
\\[4pt]
&
\Phi=\phi+F_3,
&
\label{ansatz_F3}
\eea
where $a$ and $\phi$ are background solutions depending only 
on the 5--th coordinate $y$, while the perturbations 
$F_j$ are arbitrary (but small) functions of all coordinates. 
To find the masses of the KK modes of scalar perturbations it 
is enough to consider equations of motion linear in $F_j$.

Contrary to the background equations of motion, for the 
perturbations we obtain non--trivial off--diagonal 
Einstein equations
\bea
&
2F_1+F_2=0,
\label{cond_F1F2}
\\[4pt]
&
\left(a^2 F_1\right)'+\frac13 a^2 P_X \phi' F_3=0.
&
\label{cond_F1F3}
\eea
They have to be fulfilled in order to stay in the 
longitudinal gauge. 
The diagonal Einstein equations, combined with the background
equations of motion (\ref{bulk_bg1}--\ref{bulk_bg3}), 
give the third equation for the scalar perturbations:
\bea
\Box F_1 
+ 4\frac{a'}{a}F_1' - 4 \left(\frac{a'}{a}\right)^2F_2
+\frac{a'}{a}\left(P_X-\frac23 XP_{XX}\right)\phi'F_3 
\qquad\qquad\qquad\nn\\[4pt]
+\frac13 \left(P_X+2XP_{XX}\right)
\left[{\phi'}^2 F_2 + \phi'' F_3 - \phi' F_3'\right]
=0
\,,
\label{bulk_pert}
\eea
where $\Box$ is the 4--dimensional D'Alembertian. The part of 
the boundary conditions linear in the scalar 
perturbations are quite complicated and reads 
\bea
\al
\pm2\lim_{y\to y_i^\pm}\left[\left(P_X+2XP_{XX}\right)F_3'\right]
\nn\\[4pt]
\al
+\int_{y_i}\phi''\left[\left(P_{\Phi X}+2XP_{\Phi XX}\right)F_3
-\left(P_X+8XP_{XX}+4X^2P_{XXX}\right)F_2\right]\qquad
\nn\\[4pt]
\al\qquad\qquad\qquad\qquad
=
\left.
\left[aF_3\left(
U^{(i)}_{\Phi\Phi} +\int_{y_i}\delta_i Q^{(i)}_{\Phi\Phi}\right)
-\frac{\Box F_3}{a}\int_{y_i}\delta_i Q^{(i)}_X\right]
\right|_{y=y_i}
,\qquad
\label{brane_pert}
\eea
where $\delta_i=\delta(y-y_i)$. The subscript $y_i$ at 
the integrals indicates that 
the range of integration 
is an infinitesimal interval containing $y_i$.

The off--diagonal Einstein equations (\ref{cond_F1F2}) 
and (\ref{cond_F1F3})
can be used to express two of the perturbations introduced 
in the ansatz (\ref{ansatz_F1F2}-\ref{ansatz_F3}) in terms 
of the third one. It is convenient to eliminate $F_2$ and $F_3$ 
and to use the product $a^2F_1$ as an independent perturbation.
We expand it in the 4D modes as
\be
a^2(y)F_1(t,\vec x,y)=\sum_{m^2}K_{m^2}(y)
\left[\int\d^3 k f_{(m^2,k)}(t)e^{i \vec k \vec x}\right]
,
\ee
and substitute to eqs.\ (\ref{bulk_pert}) and (\ref{brane_pert}).
Then, the 4D part of the bulk equation (\ref{bulk_pert}) takes the 
usual form
\be
\ddot f_{(m^2,k)} + \left(k^2+m^2\right)f_{(m^2,k)}=0
.
\label{bulk_f}
\ee
The equation for the ``shape'' $K_{m^2}(y)$ of the KK mode
with mass squared equal $m^2$ can be written as the Sturm--Liouville 
equation
\be
-\left(pK_{m^2}'\right)'+qK_{m^2}=m^2 r K_{m^2}
\,,
\label{bulk_K}
\ee
where $p$, $q$ and $r$ are the following functions depending on the
background
\be
p=\frac{3}{2aP_X{\phi'}^2}
\,,\qquad
q=\frac{1}{a}
\,,\qquad
r=\frac{3}{2a\left(P_X+2XP_{XX}\right){\phi'}^2}
=c_s^2 p
\,.
\ee
In the last equality we have introduced a local ($y$--dependent) 
``speed of sound''
\be
c_s^2=\frac{P_X}{P_X+2XP_{XX}}
\,.
\label{cs2}
\ee
The boundary condition (\ref{brane_pert}) in terms of $K_{m^2}$ 
takes the form
\be
\left.\left[
\left(b_i-c_im^2\right)\frac{\partial}{\partial n} K_{m^2}
-m^2 P_X K_{m^2}
\right]\right|_{y_i^\pm}=0
\,,
\label{brane_K}
\ee
where from now on $y_i^\pm$ stands 
for $y_1^+$ or $y_2^-$. The corresponding limits have
to be taken for quantities discontinuous on the branes. 
The $\partial/\partial n$ differentiation is in 
the direction of the outer normal at the boundary, i.e.\ 
$(-\d/\d y)$ at $y_1$ and $(+\d/\d y)$ at $y_2$. 
Quantities $b_i$ and $c_i$ are the following functions of
the background solution and the brane interactions 
\bea
b_i
=
\al
\frac12\left[
\left.aU^{(i)}_{\Phi\Phi}\right|_{y=y_i}
+a\int_{y_i} \delta_i Q^{(i)}_{\Phi\Phi}
-\int_{y_i}\phi''\left(P_{\Phi X}+2XP_{\Phi XX}\right)\right]
\nn\\[4pt]
\al
\mp\lim_{y\to y_i^\pm}
\left(P_X+2XP_{XX}\right)
\left(\frac{\phi''}{\phi'}-\frac{a'}{a}\right)
,
\label{b}
\\[4pt]
c_i
=
\al
\frac{1}{2a}\int_{y_i} \delta_i Q_X^{(i)}
\,.
\label{c}
\eea
All integrals in (\ref{b}) and (\ref{c}) 
should be calculated with the same 
regularization as that used in (\ref{regularization}).

The square of the radion mass is given by the lowest eigenvalue 
of the equation of motion (\ref{bulk_K}) satisfying the boundary 
conditions (\ref{brane_K}). Of course, in general it is not possible 
to find the spectrum of the system (\ref{bulk_K})-(\ref{c}) 
by solving it explicitly. To get some information about the 
smallest eigenvalue we will use methods analogous to those 
developed for a similar problem in 
\cite{KoOlSc} (where the corresponding boundary conditions 
have a form of (\ref{brane_K}) with $c_i=0$). 
But first one has to check whether the differential equation 
(\ref{bulk_K}) together with the boundary conditions 
(\ref{brane_K}) constitute a self-adjoint system.
This is a non trivial problem because conditions (\ref{brane_K})
are unusual and quite complicated.
The eigenvalue $m^2$ of the equation of motion (\ref{bulk_K}) 
appears multiplying both $K_{m^2}$ and its normal derivative.
In the next subsection we will show that our eigenvalue problem 
is self-adjoint with boundary conditions 
even more general than (\ref{brane_K}).

\subsection{Self-adjoint eigenvalue problem} 
\label{self-adjointEP}

Let us consider a differential eigenvalue problem 
\be
\cO v=\lambda v
\label{EP}
\ee
for the operator $\cO$ of the Sturm-Liouville type
\be
\cO v
=
\frac{1}{r}\left[-\left(p v'\right)'+q v\right]
\,.
\ee
The boundary conditions on the interval $(y_1,y_2)$ 
have the form 
\be
\left.\left[\sigma_1^{(i)} v + \sigma_2^{(i)} v' 
+ \sigma_3^{(i)} \left(\cO v\right) 
+ \sigma_4^{(i)} \left(\cO v\right)'\right]\right|_{y=y_i} = 0
\,,
\label{BC}
\ee
where $\sigma^{(i)}_j$ are some constants. The spectrum of our 
eigenvalue problem is real if $\cO$ is hermitian. 
In order to prove this one has to find such a scalar product 
$\left(\cdot,\cdot\right)$ for which 
\be 
\left(v,\cO u\right) = \left(\cO v, u\right)
\,.
\ee
The standard boundary conditions discussed in many mathematical 
textbooks have the form of (\ref{BC}) with 
$\sigma_3^{(i)}=\sigma_4^{(i)}=0$.
In such a case, $\cO$ is hermitian in the scalar product 
$\left(f,g\right)=\int_{y_1}^{y_2}rfg$ (for simplicity we consider 
real functions $f$ and $g$). Let us generalize this scalar product 
by adding some boundary 
terms\footnote{
A simple example of a non-standard scalar product was discussed
for example in \cite{CsHuMe}. It was calculated for a canonical 
kinetic term localized on a brane in a flat background. 
In our notation this corresponds to 
$p=1$, $q=0$, $r=1$, $\sigma_1=0$, $\sigma_4=0$.
}
\be
\left(f,g\right)=\int_{y_1}^{y_2}rfg
+\left.\left[\rho_1^{(i)} fg + \rho_2^{(i)} (fg)' 
+ \rho_3^{(i)} f' g'\right]\right|_{y_1}^{y_2}
\,,
\label{SP}
\ee
with yet unspecified constants $\rho^{(i)}_j$. 
For this scalar product we calculate
\bea
(v,\cO u) 
\al
- (\cO v, u)
=
\left\{
p\left[u v' -v u'\right]
+\rho_1^{(i)}\left[v(\cO u)-(\cO v) u\right]\right.
\nn\\[4pt]
&&
+
\rho_2^{(i)}\left[\left(v(\cO u)\right)'-\left((\cO v) u\right)'\right]
\left.\left.
+\rho_3^{(i)}\left[v' (\cO u)'- (\cO v)' u'\right]
\right\}\right|_{y_1}^{y_2}
.\qquad
\eea
Introducing three additional constants 
$\rho_4^{(i)}$, $\rho_5^{(i)}$, $p_1^{(i)}$, 
at each boundary, 
we can rewrite the above equation in the following form
\bea
(v,\cO u) 
- 
(\cO v, u)
=
\al
\left\{
v\left[\rho_4^{(i)}u-p_1^{(i)} u' + \rho_1^{(i)}(\cO u) 
+ \rho_2^{(i)} (\cO u)'
\right]\right.
\nn\\[4pt]
\al
-u\left[\rho_4^{(i)}v-p_1^{(i)} v' + \rho_1^{(i)}(\cO v) 
+ \rho_2^{(i)} (\cO v)'
\right]
\nn\\[4pt]
\al
+v'\left[p_2^{(i)} u + \rho_5^{(i)} u' + \rho_2^{(i)}(\cO u) 
+ \rho_3^{(i)} (\cO u)'
\right]
\nn\\
\al
-\left.\left.\!\!
u'\left[p_2^{(i)} v + \rho_5^{(i)} v' + \rho_2^{(i)}(\cO v) 
+ \rho_3^{(i)} (\cO v)'
\right]\right\}\right|_{y_1}^{y_2}
\qquad
\label{herm2}
\eea
where $p_2^{(i)}=p(y_i)-p_1^{(i)}$. 
Our operator $\cO$ is hermitian if the r.h.s.\ of the above
equation vanishes for all $v$ and $u$ fulfilling 
the boundary conditions (\ref{BC}). 
This is the case when each square bracket in (\ref{herm2}) 
is proportional the square bracket in (\ref{BC}): 
\bea
\rho_4^{(i)}=n_1^{(i)}\sigma_1^{(i)}
\,,\quad
-p_1^{(i)}=n_1^{(i)}\sigma_2^{(i)}
\,,\quad
\rho_1^{(i)}=n_1^{(i)}\sigma_3^{(i)}
\,,\quad
\rho_2^{(i)}=n_1^{(i)}\sigma_4^{(i)}
\,,
\\[4pt]
p_2^{(i)}=n_2^{(i)}\sigma_1^{(i)}
\,,\quad
\rho_5^{(i)}=n_2^{(i)}\sigma_2^{(i)}
\,,\quad
\rho_2^{(i)}=n_2^{(i)}\sigma_3^{(i)}
\,,\quad
\rho_3^{(i)}=n_2^{(i)}\sigma_4^{(i)}
\,.
\eea
For generic values of $\sigma_j^{(i)}$ this set of 
linear equations can be easily solved. At each 
boundary there are 8 equations and 8 independent constants:
$\rho_1^{(i)}$, $\rho_2^{(i)}$, $\rho_3^{(i)}$, $\rho_4^{(i)}$, 
$\rho_5^{(i)}$, $n_1^{(i)}$, $n_2^{(i)}$, $p_1^{(i)}$. 
In fact we are interested only in those three, 
$\rho_1^{(i)}$, $\rho_2^{(i)}$, $\rho_3^{(i)}$, 
which enter the definition of the scalar product (\ref{SP}). 
The solution 
reads
\bea
\left(f,g\right)=\int_{y_1}^{y_2}rfg
\left.
+\left[p\,
\frac{
\left(\sigma_3^{(i)}\right)^2 fg 
+ \sigma_3^{(i)}\sigma_4^{(i)} (fg)' 
+ \left(\sigma_4^{(i)}\right)^2 f' g'}
{\sigma_1^{(i)}\sigma_4^{(i)}-\sigma_2^{(i)}\sigma_3^{(i)}}
\right]
\right|_{y_1}^{y_2}
\!\!.\,\,
\label{SP_sol}
\eea 
We have shown that the eigenvalue problem (\ref{EP}) with 
the boundary conditions (\ref{BC}) is self-adjoint. 
Thus, all its eigenvalues $\lambda$ are real and the eigenfunctions 
corresponding to different $\lambda$ are orthogonal in the 
scalar product (\ref{SP_sol}).

Let us now use the above result for our k-stabilization mechanism.
The boundary conditions (\ref{brane_K}) have the form of  
(\ref{BC}) with
\be
\sigma_1^{(i)}=0
\,,\quad
\sigma_2^{(i)}=(-1)^i b_i
\,,\quad
\sigma_3^{(i)}=-P_X(y_i)
\,,\quad
\sigma_4^{(i)}=-(-1)^i c_i
\,,
\label{sigma_k}
\ee
with no summation over $i$.
The factors of $(-1)^i$ appear because the outer normal 
derivative $\partial/\partial n$ was used in (\ref{brane_K}). 
Substituting (\ref{sigma_k}) into (\ref{SP_sol}) we obtain 
the following scalar product appropriate to show that the eigenvalue 
problem (\ref{bulk_K}), (\ref{brane_K}) is self-adjoint:
\be
\left(f,g\right)=\int_{y_1}^{y_2}rfg
+{\sum_{i=1,2}}'
\left.\left[ p\,
\frac{
P_X^2 fg 
+ P_X c_i \ddn (fg) 
+ c_i^2 \ddn f \ddn g}
{P_X b_i}
\right]
\right|_{y_i}
.
\ee
The prime at the sum symbol denotes that 
the boundary contributions should be taken only for those 
boundaries at which $P_Xb_i\ne0$. The reason is that for 
$b_i=0$ and/or $P_X=0$ the boundary condition (\ref{brane_K}) 
reduces to the standard one for which 
$\left(f,g\right)=\int rfg$ without any boundary 
terms (at that boundary).

\section{Stability conditions}
\label{stability}

The spectrum of the scalar perturbations in a given background 
is given by the eigenvalues of the Strum--Liouville equation 
(\ref{bulk_K}) with the boundary conditions (\ref{brane_K}) 
at the branes. 
In the previous subsection we have shown that 
this spectrum is real.
The most interesting for us is the lowest 
eigenvalue which we identify with the square of the radion mass.
The inter--brane distance is stable only if this mass squared 
is positive. In this section we will find conditions sufficient 
for such stability. We will show also when the radion is massless 
and identify some classes of backgrounds which are unstable.

First we check whether there is a massless mode in the KK tower 
of the scalar perturbations. In such a case, the
boundary condition (\ref{brane_K}) at the first brane 
reduces, for nonzero $b_1$, to $K'_0(y_1^+)=0$
(the case with vanishing $b_1$ will be considered later).
For $m^2=0$, the solution of the bulk equation of motion 
(\ref{bulk_K}), satisfying 
the boundary condition at $y=y_1$ and normalized to
$K_0(y_1)=1$, can be written in quite a simple form
\be
K_0(y)
=
\frac{a^2(y)}{a^2(y_1)}
-\frac{2a'(y)}{a^2(y)a^2(y_1)}
\int_{y_1}^y\d {\tilde y}\,a^3({\tilde y})
\,.
\label{K0}
\ee
Using the background equation of motion (\ref{bulk_bg2}), 
the derivative of the above solution simplifies to
\be
K'_0(y)
=
\frac{2P_X(y) {\phi'}^2(y)}{3a(y)a^2(y_1)}
\int_{y_1}^y\d {\tilde y}\,a^3({\tilde y})
\,.
\label{K0'}
\ee
The boundary condition at the second brane reads 
$b_2K'(y_2^-)=0$. The integral in eq.\ 
(\ref{K0'}) is strictly positive, so this condition 
is fulfilled only when 
the product $b_2 P_X(y_2^-) \phi'(y_2^-)$ vanishes. 
Repeating the same reasoning starting from the second brane, 
one gets analogous result for the first brane. Putting 
both cases together, we find that for $p$ and $r$ regular 
in the bulk, the necessary and sufficient condition 
for existence of a massless mode is
\be
b_1b_2P_X(y_1^+) P_X(y_2^-) \phi'(y_1^+) \phi'(y_2^-)=0
\,.
\ee

Conditions sufficient for the stability can be
found in the following way. 
Multiplying eq.\ (\ref{bulk_K}) with $K_{m^2}$, integrating 
over the whole 5th dimension, and using the boundary 
conditions (\ref{brane_K}) we get
\bea
m^2\int_{y_1}^{y_2} r(K_{m^2})^2
\al
+\left.\sum_i\frac{b_i}{m^2}\frac{p}{P_X}(K'_{m^2})^2\right|_{y_i^\pm}
\nn\\[4pt]
\al=
\int_{y_1}^{y_2} \left[q(K_{m^2})^2+p(K'_{m^2})^2\right]
+\left.\sum_ic_i\frac{p}{P_X}(K'_{m^2})^2\right|_{y_i^\pm}
.\quad
\label{m2}
\eea
Let us consider first such models for which 
the background dependent 
bulk functions $p$, $q$ and $r$ are regular and positive 
while the brane parameters $b_i$ are positive and $c_i$ 
are non--negative. Then, the r.h.s.\ of (\ref{m2}) is 
positive while the l.h.s.\ is negative for negative $m^2$ 
and may be divergent for vanishing $m^2$. Thus, 
the condition (\ref{m2}) can be fulfilled only 
for positive $m^2$. The function 
$q=1/a$ is always positive. Functions $p$ and $r$ have 
the same sign as $P_X$ and $(P_X+2XP_{XX})$, respectively. 
They become singular if any of the functions 
$P_X$, $(P_X+2XP_{XX})$ or $\phi'$ 
vanishes for any value of $y$. 
Thus, the following conditions 
\bea
&
b_i>0,
\qquad\qquad
c_i\ge0,
\label{stability_b}
&
\\[4pt]
&
\forall_{y\in[y_1^+,\,y_2^-]}
\qquad
{\phi'}^2(y)>0,
\quad
P_X(y)>0,
\quad
P_X(y)+2X(y)P_{XX}(y)>0
,\quad
&
\label{stability_B}
\eea
are sufficient for the stability of the 
inter--brane distance (positivity of the radion mass  
squared). By $y\in[y_1^+,y_2^-]$ we denote the interior 
of the bulk, $y_1<y<y_2$ and the limits $y\to y_1^+$ 
and $y\to y_2^-$.

Showing that the above conditions are sufficient for 
stability was quite easy.
It is much more difficult to check which conditions are
necessary.
We will show now that there must be at least one tachyonic mode 
if any of the functions, $\phi'$, $P_X$ or $(P_X+2XP_{XX})$, 
vanishes anywhere in the bulk. 
The arguments are similar to those used in 
\cite{LeSo} and \cite{KoOlSc}. We will compare the properties 
of two solutions of the bulk equation of motion (\ref{bulk_K}),
one for $m^2=0$ and second for $m^2=-M^2$ in the limit $M\to\infty$.
Both solutions satisfy the boundary condition at 
one brane (let us first choose it to be the first one 
located at $y_1$).

We start with 
solving the bulk equation of motion (\ref{bulk_K}) in the 
limit of large negative $m^2=-M^2$. 
In the leading order in $1/M$, equation (\ref{bulk_K}) 
has the following approximate solution
\bea
K_{-M^2}(y)
\approx\al
\frac{1}{\sqrt{pc_s}}
\left[C^+\exp\left(+M\int_{y_1}^yc_s\right)
+C^-\exp\left(-M\int_{y_1}^yc_s\right)\right]
,
\label{Kinf_pm}
\\[4pt]
K'_{-M^2}(y)
\approx\al
M\sqrt{\frac{c_s}{p}}
\left[C^+\exp\left(+M\int_{y_1}^yc_s\right)
-C^-\exp\left(-M\int_{y_1}^yc_s\right)\right]
.\quad
\label{K'inf_pm}
\eea
In the same limit, the boundary condition (\ref{brane_K}) 
at the first brane becomes
\be
\left.\left(c_1K_{-M^2}'-P_XK_{-M^2}\right)\right|_{y=y_1}\approx0
\,.
\label{Kinf_y1}
\ee
Because of the $M$ prefactor in (\ref{K'inf_pm}), for any $c_1\ne0$ 
and large enough $M$, the above boundary condition can be fulfilled 
when $C^+\approx C^-$.
We choose $C^\pm$ to be positive because later we will  
compare this solution with $K_0$ normalized to 1 at $y_1$.
When $c_1$ and $P_X(y_1)$ have the same sign, the boundary condition 
(\ref{Kinf_y1}) can be fulfilled only when $K_{-M^2}(y_1)$ 
and $K'_{-M^2}(y_1)$ have the same sign. Thus,  
$C^+>C_-$ and the square bracket in (\ref{Kinf_pm}) does 
not change its sign in the whole bulk.
For very large $M$ the first term in 
(\ref{Kinf_pm}) starts do dominate over the second one even for 
small values of $y-y_1$ (it is slightly bigger even at $y_1$) 
and away from the first brane the solution is approximated by
\be
K_{-M^2}(y)\approx{C^+}\phi'\sqrt{\frac{2a}{3}
\sqrt{P_X\left(P_X+2XP_{XX}\right)}}
\exp\left(M\int\frac{P_X}{P_X+2XP_{XX}}\right)
.
\label{Kinf}
\ee
Using this solution we can investigate models when 
some of the conditions in 
(\ref{stability_b}-\ref{stability_B}) are not fulfilled.

It is convenient to define the following function of $m^2$
\be
B_2(m^2)
=
\left.\left[
\left(b_2-c_2 m^2\right)\frac{\partial}{\partial n} K_{m^2}
-m^2 P_X K_{m^2}
\right]\right|_{y=y_2^-}.
\label{B2}
\ee
It is equal to the l.h.s.\ of the boundary 
condition (\ref{brane_K}) 
for $K_{m^2}$ satisfying the bulk equation of motion 
and the boundary condition at the first brane, and normalized 
to 1 at $y_1$. 
The spectrum of the KK tower of scalar perturbations 
consists of those values $m^2$ for which $B_2(m^2)=0$.

Now we check whether the positivity of $b_i$ and $c_i$ 
are necessary conditions for the stability, assuming that
all the bulk conditions (\ref{stability_B}) are fulfilled.
For very lage negative $m^2$ the boundary function 
$B_2$ at the second brane is dominated 
by the term proportional to $K'_{-M^2}$. From eq.\ (\ref{Kinf}) 
and the discussion before it, it follows that
\be
\sgn\left[B_2(-M^2)\right]
=
\sgn\left[M^2 c_2 K'_{-M^2}(y_2^-)\right]
=
\sgn\left[c_2\right]
\,.
\label{B2_-M2}
\ee
On the other hand, for the solution $K_0$ given by 
(\ref{K0}) and (\ref{K0'}) we get
\be
\sgn\left[B_2(0)\right]
=
\sgn\left[b_2K'_0(y_2^-)\right]=\sgn[b_2]
\label{B2_0}
\,,
\ee
where we used the fact that $K'_0$ is always positive 
when the inequalities (\ref{stability_B}) are fulfilled.
Comparing (\ref{B2_-M2}) with (\ref{B2_0}), we conclude 
that there must be at least one negative eigenvalue 
when the parameters $b_2$ and $c_2$ have opposite signs. 
For $b_2 c_2<0$, the function $B_2(m^2)$ has different 
sign for $m^2=0$ and for large (enough) negative $m^2$. 
There must be some negative $m^2$ for which $B_2$ 
vanishes because the solutions of (\ref{bulk_K}) change 
continuously with $m^2$.

Repeating the above reasoning but starting from the brane 
at $y_2$, we obtain an analogous condition for parameters 
$b_1$ and $c_1$. Thus, the conditions
\be
b_1c_1\ge0\,,
\qquad\qquad
b_2c_2\ge0\,,
\label{stability_bc0}
\ee
are necessary for the stability.

Now we investigate the stability conditions 
for the bulk quantities (\ref{stability_B}).
The solution (\ref{Kinf}) for large negative $m^2$ 
vanishes at a point at which $\phi'$ or $P_X$ vanishes. 
It must change sign there because from (\ref{bulk_K}) 
it follows that $K$ and $K'$ can vanish at the same point 
only for trivial solution vanishing everywhere.
Thus, for very large negative $m^2$ the function $K(y)$
vanishes close to the point where $P_X\phi'$ is zero.
On the other hand, from (\ref{K0}) and (\ref{K0'}) it follows 
that $K_0$ is positive for all $y$. So, there must be 
some negative $\widetilde m^2$ for which $K_{\widetilde m^2}$ 
has a zero point but is nowhere negative. It is easy to 
see that such a zero point must be at the second brane,
$y=y_2$, and that the derivative of $K_{\widetilde m^2}(y_2^-)$ 
is negative. In such a situation 
\be
\sgn\left[B_2(\widetilde m^2)\right]
=
\sgn\left[(b_2-c_2 \widetilde m^2) 
K'_{\widetilde m^2}(y_2^-)\right]
=-\sgn[b_2]
\,,
\label{B2_-m2}
\ee
where the last equality follows from the condition 
(\ref{stability_bc0}). 
Comparing (\ref{B2_0}) and (\ref{B2_-m2}) we find that there must be 
some negative mode with the eigenvalue ${\widehat m}^2$ 
satisfying $\widetilde m^2<{\widehat m}^2<0$ for which 
$B_2({\widehat m}^2)=0$. The radion is tachyonic if 
$\phi'$ or $P_X$ vanishes in the bulk.

The above arguments are rather complicated but the 
result is quite intuitive. We consider backgrounds for 
which $P_X\phi'$ vanishes at some $y_0<y_2$ in the bulk. 
For any such background $K_0(y)$ defined in (\ref{K0}) 
is a zero mode in a model restricted to the interval 
$[y_1,y_0]$. It is quite natural that the KK states becomes 
lighter when the compact space becomes bigger. So, with 
a massless mode on $[y_1,y_0]$ there should be a tachyonic 
one on the bigger orbifold $[y_1,y_2]$.

Equation (\ref{Kinf}) can be used to show that also 
$(P_X+2XP_{XX})$ should be strictly positive. 
If it is not, there are two possibilities 
depending on how fast it approaches zero. If the integral 
in (\ref{Kinf}) is finite then $K_{-M^2}$ vanishes 
because of $(P_X+2XP_{XX})$ in the prefactor and a 
reasoning similar to that for the case of vanishing 
$P_X\phi'$ may be 
applied to prove the existence of at least one tachyonic mode.
On the other hand, a divergent integral in (\ref{Kinf}) 
indicates the breakdown of the perturbativity  
assumption. This is not surprising. Vanishing $(P_X+2XP_{XX})$ 
corresponds to infinite speed of sound while negative 
$(P_X+2XP_{XX})$ gives negative square of the speed of sound 
(for positive $P_X$, which is anyway necessary for the stability).
In both cases one should expect strong instabilities.

We showed above that the conditions (\ref{stability_B}) 
on the bulk quantities are not only sufficient but 
also necessary for the stability. 
We were not able to prove the same for the brane conditions 
(\ref{stability_b}). If one of them is fulfilled then the other
has also to be fulfilled. The only possible loophole occurs 
when both conditions (\ref{stability_b}) are violate,
namely when $b_1<0$ and $c_1<0$ or when $b_2<0$ and $c_2<0$.
However, these are not very appealing possibilities. 
Parameters $c_i$ are proportional to the integrals 
$\int_{y_i}\delta_iQ^{(i)}_X$ and can be negative only 
for localized brane kinetic terms very different from
the standard one.

\section{Discussion}
\label{discussion}

With the results presented in the two previous sections 
we can investigate how the stabilization of 
branes is influenced by the presence of 
non--trivial scalar kinetic terms in the bulk and/or 
on the branes. Such terms change the background 
configurations and the spectrum of the scalar perturbations.
We start the discussion with the background.

Combining eqs.\ (\ref{bulk_bg2}) 
and (\ref{bulk_bg3}), the dynamical equation describing 
the change of the warp factor can be written as 
\be
3\frac{a''}{a}+a^2\left(V+P\right)=0\,.
\ee
The source for the change of the warp 
factor $a(y)$ is the full ``matter'' Lagrangian density 
$(V+P)$ 
irrespective of whether the kinetic part is standard 
or not. The modification of the scalar equation of motion 
given in (\ref{bulk_bg1}) 
\be
\left(P_X\phi'\right)'+3\frac{a'}{a}\left(P_X\phi'\right)
-a^2\left(V_\Phi+P_\Phi\right)=0\,,
\nn
\ee
is twofold. First, similarly as in the case of the 
warp factor, the role of the potential in this equation 
is played by the full non--gravitational Lagrangian density. 
Second, it seems that a natural 
variable to describe the change of the scalar 
background is the product $P_X\phi'$ 
and not $\phi'$ itself. The equation of motion 
for this generalized variable $P_X\phi'$ looks formally 
the same as that in the standard theory (derivative 
of the full Lagrangian as a source and $3a'/a$ as ``friction''). 
Thus, as compared to the standard theory, 
for the same local non--gravitational energy density 
and the warp factor slope, the scalar 
field $\phi$ changes faster (slower) if $P_X$ is smaller 
(bigger) than 1. Of course this is only a qualitative 
feature and in most of the cases any quantitative corrections 
can be found only by numerical calculations.

The positions of the branes are determined by 
the boundary conditions. The modifications to the boundary 
conditions (\ref{brane_a}) and (\ref{brane_phi}) are 
analogous to those in the bulk background equations.
Namely, not only the potentials but the full Lagrangians 
localized at the branes determine the jumps of $a'$ 
while their derivatives with respect to $\Phi$ determine 
the jumps of $P_X\phi'$.

Usually in Randall--Sundrum type models, the warp factor changes 
monotonically in the bulk, so its derivative has the same 
sign for all $y$. Thus, because of opposite overall signs 
in the boundary conditions (\ref{brane_a}) at two branes, one brane 
must have positive tension while the second one must have 
negative tension. To check the signs of the brane tensions 
in the class of models considered in this work we rewrite 
eq.\ (\ref{bulk_bg2}) in the following form
\be
\left(\frac{a'}{a^2}\right)'=-\frac{1}{3a}P_X{\phi'}^2
\,.
\ee
In the previous section we showed that the stability of the 
model requires that $P_X\phi'$ is everywhere non--zero. Thus, 
the r.h.s.\ of the above equation is always negative. 
The ratio $a'/a^2$ always decreases and the warp factor $a(y)$ 
can not have a minimum in the bulk. Because of that, 
it is not possible to construct a stable model with two 
positive tension branes. At least 
one brane must have a negative tension: 
\be
\min_i\left(
\left.U^{(i)}\right|_{y_i} + \int_{y_i}\delta_iQ^{(i)}
\right)
<0
\,.
\label{U1U2}
\ee
In all stable models $\phi(y)$ must be a monotonic function
($\phi'$ can not vanish) and $P_X$ can not change sign.
The limit of the product $P_X\phi'$ has the same sign at 
both branes. Thus, it follows from the boundary 
condition (\ref{brane_phi}) that
\be
\left(\left.U^{(1)}_\Phi\right|_{y_1} 
+ \int_{y_1}\delta_1Q^{(1)}_\Phi\right)
\cdot
\left(\left.U^{(2)}_\Phi\right|_{y_2} 
+ \int_{y_2}\delta_2Q^{(2)}_\Phi\right)
<0
\,.
\label{U1'U2'}
\ee

We turn now to the stability conditions. One of them is 
the positivity of $b_i$ parameters defined in (\ref{brane_K}). 
Using the background equation of motion (\ref{bulk_bg1}) 
the last term in the definition of $b_i$ 
(\ref{b}) can be rewritten as
\bea
\al
\mp\lim_{y\to y_i^\pm}
\left(P_X+2XP_{XX}\right)
\left(\frac{\phi''}{\phi'}-\frac{a'}{a}\right)
\nn\\[4pt]
\al\qquad\qquad
=
\lim_{y\to y_i^\pm}
\left[
\pm 4 P_X \frac{a'}{a}
\pm P_{\Phi X}\phi'
\mp \frac{a^2\left(V_{\Phi}+P_{\Phi}\right)}{\phi'}
\right]
\nn\\[4pt]
\al\qquad\qquad
=
\lim_{y\to y_i^\pm}
\left[
- 4 P_X \frac{\frac{\partial}{\partial n}a}{a}
- P_{\Phi X}\frac{\partial}{\partial n}\phi
+ \frac{a^2\left(V_{\Phi}+P_{\Phi}\right)}
{\frac{\partial}{\partial n}\phi}
\right]
,
\label{bi_last}
\eea
where we used the outer normal derivative introduced in 
eq.\ (\ref{brane_K}). The first term in the last square bracket  
of the above equation gives negative (positive) contribution 
to the $b$ parameter on the positive (negative) tension brane. 
So, positivity of $b$ at the positive tension brane 
is more difficult to achieve. Stability is improved when, 
close to the brane(s), $P_{\Phi X}$ has the opposite sign and
$\left(V_{\Phi}+P_{\Phi}\right)$ has the same sign as the 
normal derivative of the scalar field $\partial\phi/\partial n$.
There is another term in the definition of $b$ which depends 
on the bulk background, namely 
$\int\phi''(P_{\Phi X}+2XP_{\Phi XX})$. Its sign depends 
on the background and on the details of the generalized 
bulk kinetic function $P$. Non--trivial $\Phi$--dependence 
of $P$ can be, at least in some cases, used to increase 
the radion mass. Finally, large enough values of the second 
derivatives of the brane kinetic terms $Q^{(i)}$ may be 
used to make $b_i$ positive.

The second stability condition in (\ref{stability_b}) 
can be quite easily 
fulfilled. For example: $c_i$ given by eq.\ (\ref{c}) 
vanishes if there is no kinetic 
term localized on the $i$--th brane and it is positive when 
such localized term is not much different from the standard 
one $Q^{(i)}=X$.

Models with non--standard bulk and/or brane scalar kinetic 
terms are quite complicated and usually only performing numerical 
calculations one can find the background fields and check 
their stability against small perturbations. 
Nevertheless, it seems viable that stable 
solutions can exist also in models without any scalar 
potentials or cosmological constants. The kinetic terms alone 
may have structure rich enough for 
configurations with stabilized inter--brane distance.
This is similar to the situation in models 
proposed in \cite{ArDaMu,GaMu} in which inflation was 
realized without any scalar potential.

Let us discuss what properties the generalized kinetic 
terms should have in order to support stable brane 
configurations. Conditions on the bulk kinetic function 
$P$ are rather weak. It is enough that eq.\ (\ref{bulk_bg3}) 
can be fulfilled for some $y_0$ and positive values of 
${\phi'}^2$, $P_X$ and $(P_X+2XP_{XX})$. Then, the dynamical 
equations (\ref{bulk_bg1}) and (\ref{bulk_bg2}) can be used to 
extend the solution to $y\ne y_0$. The bulk stability conditions 
(\ref{stability_B}) are fulfilled at $y_0$, so they are 
fulfilled also in some finite interval in the 5th coordinate.
Any two points in this interval may be used to locate 
the branes. Of course, this is possible only when the brane 
kinetic terms have appropriate properties.

Restrictions on the brane kinetic functions $Q^{(i)}$ 
are quite strong 
if we want the branes to be stabilized 
at given positions in a given background. 
First of all, from eq.\ (\ref{U1'U2'}) it is obvious that 
without brane potentials it is necessary 
that $Q^{(i)}$ have some non--trivial 
$\Phi$--dependence. In addition, it follows 
from (\ref{U1U2}) that at least at one of the branes 
the kinetic term must give a negative contribution to its  
tension.
This does not a priori mean that the system becomes 
unstable. Of course, we want the energy to be bounded from 
below, so the kinetic term at the second brane 
$Q^{(2)}(\Phi,X)$ (we call ``second'' that brane at which 
the expression in (\ref{U1U2}) is minimized) 
should give negative value of $\int\delta_2Q^{(2)}$ 
only for some range of values of its 
arguments\footnote
{
One should remember that in general $Q^{(i)}$ and 
$\int\delta_iQ^{(i)}$ 
are not just proportional to 
each other and can have negative values for 
(slightly) different regions of the parameter space.
This is caused by the regularization procedure discussed 
in section \ref{model}.
}.  
More specifically, each $Q^{(i)}$ must satisfy 
two equalities (\ref{brane_a}) 
and (\ref{brane_phi}) and two inequalities (\ref{stability_b}).
The values of $\int\delta_iQ^{(i)}$ and 
$\int\delta_iQ^{(i)}_\Phi$ necessary to fulfill the 
background boundary 
conditions depend on the details of a given background 
but their signs are determined by (\ref{U1U2}) and (\ref{U1'U2'}). 
All boundary and stability conditions on the brane kinetic 
functions may be written in the following form
\bea
\int_{y_i}\delta_i Q^{(i)}
\al=
\left.\frac{6}{a^2}\frac{\partial a}{\partial n}\right|_{y_i^\pm}
\,,
\label{Qa}
\\[4pt]
\int_{y_i}\delta_i Q^{(i)}_\Phi
\al=
-\left.\frac{2P_X}{a}\frac{\partial\phi}{\partial n}\right|_{y_i^\pm}
\,,
\label{Qphi}
\\[4pt]
\int_{y_i}\delta_i Q^{(i)}_X
\al\ge
0
\,,
\label{Qc}
\\[4pt]
\int_{y_i}\delta_i Q^{(i)}_{\Phi\Phi}
\al>
-\frac{2\tilde b_i}{a(y_i)}
\,,
\label{Qb}
\eea
where $\tilde b_i$ is the r.h.s.\ of (\ref{b}) with
$Q^{(i)}$ set to zero. It is possible to fulfill all the 
above conditions for example with the brane kinetic functions 
of the form 
\be
Q^{(i)}=K^{(i)}(\Phi)X+L^{(i)}(\Phi)X^2
\,.
\label{QKL}
\ee
The most difficult part is to satisfy simultaneously conditions 
(\ref{Qa}) and (\ref{Qc}) at the second (negative tension) brane. 
Using eqs.\ (\ref{Qa}), (\ref{Qc}) and (\ref{regularization}), 
one can find a lower bounds on $L^{(i)}$
\be
\left.L^{(i)}X^2
\right|_{y_i^\pm}
\ge
\left.-\frac{270}{a^2}\,\frac{\partial a}{\partial n}
\right|_{y_i^\pm}
\,,
\ee
which can be translated to an upper bound on $K^{(i)}$
\be
\left.K^{(i)}X
\right|_{y_i^\pm}
=
-\left.\frac35 L^{(i)}X^2
\right|_{y_i^\pm}
+\left.\frac{18}{a^2}\frac{\partial a}{\partial n}
\right|_{y_i^\pm}
\le
\left.\frac{180}{a^2}\,\frac{\partial a}{\partial n}
\right|_{y_i^\pm}
\,.
\ee
At the positive tension brane $L^{(2)}(\phi(y_2))$ must be 
positive and big enough while $K^{(2)}(\phi(y_2))$ must be negative
(with value related to the value of $L^{(2)}$).
Thus, without scalar potentials it is not possible to construct 
a stable model with a positive tension brane if 
the corresponding $K^{(i)}$ is always positive.

Some $\Phi$--dependence of $K^{(i)}$ and/or $L^{(i)}$ is necessary to 
fulfill conditions (\ref{Qphi}) and (\ref{Qb}). The background 
boundary condition (\ref{Qphi}) takes the following form
\be
\left.\left[
\frac13 K^{(i)}_\Phi X +\frac15 L^{(i)}_\Phi X^2\right]\right|_{y_i^\pm}
=
-\left.\frac{2P_X}{a}\frac{\partial\phi}{\partial n}\right|_{y_i^\pm}
\,.
\ee
In all stable configurations, the r.h.s.\ of this equation has 
opposite signs on the two branes (because $\phi'$ can not change 
sign). So, there are no consistent brane models without potentials 
if all first derivatives of $K^{(i)}$ and $L^{(i)}$ have the same sign.

The stability conditions (\ref{Qb}) for the brane kinetic functions 
(\ref{QKL}) may be rewritten as
\bea
\lim_{y\to y_i^{\pm}}
\al
\left[\frac13K^{(i)}_{\Phi\Phi}X+\frac15L^{(i)}_{\Phi\Phi}X^2
+
\frac{P_{\Phi X}}{P_X}
\left(\frac13K^{(i)}_{\Phi}X+\frac15L^{(i)}_{\Phi}X^2\right)
\right.
\nn\\[4pt]
\al\,\,\,\,
-\left.4P_X
\left(\frac19K^{(i)}X+\frac{1}{15}L^{(i)}X^2
+\frac{P_\Phi}{\frac13K^{(i)}_\Phi X+\frac15L^{(i)}_\Phi X^2}\right)
\right]
\nn\\[4pt]
\al\qquad\qquad\qquad\qquad\qquad\qquad\qquad
>\int_{y_i}\frac{\phi''\left(P_{\Phi X}+2XP_{\Phi XX}\right)}{a}
\,.\quad
\eea
Some of the terms on the l.h.s.\ of the the above expression 
may be negative but they can be compensated 
by large enough value of $K_{\Phi\Phi}^{(i)}X/3+L_{\Phi\Phi}^{(i)}X^2/5$.

It is clear that it is possible to choose functions 
$K^{(i)}$ and $L^{(i)}$ which satisfy all the above boundary 
and stability conditions for a given background. So, 
models in which the inter--brane distance is fixed in a stable 
way can be constructed even without any scalar potentials 
or cosmological constants. 
The brane induced kinetic terms may have 
a relatively simple form $Q=KX+LX^2$ if the functions 
$K$ and $L$ are generic 
enough\footnote
{Of course, one fine tuning of parameters is necessary 
as in all models with flat 4D foliation.
}. 
It would be interesting 
to check whether any higher order kinetic terms predicted 
for example by string theories have an appropriate structure.

\section{Conclusions}
\label{conclusions}

We considered 5D brane models with bulk and 
brane scalar kinetic terms generalized to 
some functions of $X=(\nabla\phi)^2/2$ and the scalar 
field itself. The background equations of 
motion and boundary conditions have structure similar to 
the case with standard kinetic terms. There are two kinds 
of modifications. First: the scalar potential is replaced by the 
sum of the potential and the kinetic term. Second: derivatives 
of the scalar field are multiplied by derivatives of the bulk 
kinetic term with respect to $X$.

Stability of background configurations has been checked 
by analyzing the spectrum of small scalar perturbations.
A given background with fixed branes positions 
is stable only when all 
the masses squared in the spectrum are positive. 
The bulk equation of motion determining the shape of 
the KK modes of such perturbations was written in 
the Sturm--Liouville form.  
The corresponding boundary conditions have rather 
complicated form. They may be expressed in terms of 
four parameters 
(two for each brane), $b_i$ and $c_i$, 
determined by the background and by the bulk interactions 
described effectively by some potentials and generalized 
kinetic terms. The boundary conditions depend 
also on the eigenvalues and this dependence is 
more complicated than in models with standard kinetic terms. 
We have shown that our eigenvalue problem is self-adjoint  
with those complicated boundary conditions. 
We identified even larger class of boundary conditions 
for which the Sturm-Liouville operator is hermitian.

The eigenvalue--dependence of the boundary conditions 
makes the stability considerations more difficult. 
Sufficient conditions for the stability are:
$b_i>0$, $c_i\ge 0$ at each brane 
and the positivity of bulk functions 
$P_X$, $(P_X+2XP_{XX})$ and $\phi'^2$ 
for all values of the 5th 
coordinate $y$. If $c_i\ge 0$ then the remaining conditions 
are not only sufficient but also the necessary ones. 
This changes when any of the $c_i$ parameters is 
negative. It seems that it may be possible to have stable 
configurations with negative both $b_1$ and $c_1$ (or $b_2$ and $c_2$).
The lowest KK mode, the radion, becomes tachyonic 
when any of the quantities $b_ic_i<0$ or any of the quantities 
$\phi'^2$, $P_X$ or $(P_X+2XP_{XX})$ is not strictly positive.

We have shown that stable brane models may be constructed 
without bulk and/or brane potentials and cosmological 
constants. This may be achieved for example when 
the brane localized kinetic terms take the form 
$Q^{(i)}=K^{(i)}(\Phi)X+L^{(i)}(\Phi)X^2$. 
Conditions for the functions $K^{(i)}(\Phi)$ and 
$L^{(i)}(\Phi)$ have been found.

\section*{Acknowledgments}
This work has been supported by a Marie Curie Transfer of Knowledge
Fellowship of the European Community's Sixth Framework Programme 
under contract number MTKD-CT-2005-029466 (2006-2010).
The author would like to thank for the hospitality experienced at 
Ludwig Maximilian University and Max Planck Institute in Munich 
where this work has been done.



\begin{thebibliography}{99}

\bibitem{GoWi}
W.~D.~Goldberger and M.~B.~Wise,
Phys.\ Rev.\ Lett.\  {\bf 83} (1999) 4922
[arXiv:hep-ph/9907447].

\bibitem{ChGrRu}
C.~Charmousis, R.~Gregory and V.~A.~Rubakov,
Phys.\ Rev.\ D {\bf 62} (2000) 067505
[arXiv:hep-th/9912160].

\bibitem{DWFrGuKa}
  O.~DeWolfe, D.~Z.~Freedman, S.~S.~Gubser and A.~Karch,
  Phys.\ Rev.\  D {\bf 62}, 046008 (2000)
  [arXiv:hep-th/9909134].

\bibitem{TaMo}
T.~Tanaka and X.~Montes,
Nucl.\ Phys.\ B {\bf 582} (2000) 259
[arXiv:hep-th/0001092].

\bibitem{CsGrKr}
C.~Cs\'aki, M.~L.~Graesser and G.~D.~Kribs,
Phys.\ Rev.\ D {\bf 63} (2001) 065002
[arXiv:hep-th/0008151].

\bibitem{MuKo}
  S.~Mukohyama and L.~Kofman,
  Phys.\ Rev.\  D {\bf 65}, 124025 (2002)
  [arXiv:hep-th/0112115].

\bibitem{LeSo}
J.~Lesgourgues and L.~Sorbo,
Phys.\ Rev.\ D {\bf 69} (2004) 084010
[arXiv:hep-th/0310007].

\bibitem{FrKo}
A.~V.~Frolov and L.~Kofman,
Phys.\ Rev.\ D {\bf 69} (2004) 044021
[arXiv:hep-th/0309002].

\bibitem{KoOlSc}
 D.~Konikowska, M.~Olechowski and M.~G.~Schmidt,
 Phys.\ Rev.\  D {\bf 73}, 105018 (2006)
 [arXiv:hep-th/0603014].

\bibitem{ArDaMu}
C.~Armendariz-Picon, T.~Damour and V.~F.~Mukhanov,
  Phys.\ Lett.\  B {\bf 458}, 209 (1999)
  [arXiv:hep-th/9904075].

\bibitem{GaMu}
J.~Garriga and V.~F.~Mukhanov,
  Phys.\ Lett.\  B {\bf 458}, 219 (1999)
  [arXiv:hep-th/9904176].

\bibitem{ChOkYa}
  T.~Chiba, T.~Okabe and M.~Yamaguchi,
  Phys.\ Rev.\  D {\bf 62}, 023511 (2000)
  [arXiv:astro-ph/9912463].

\bibitem{ArMuSt}
C.~Armendariz-Picon, V.~F.~Mukhanov and P.~J.~Steinhardt,
  Phys.\ Rev.\ Lett.\  {\bf 85}, 4438 (2000)
  [arXiv:astro-ph/0004134];
  Phys.\ Rev.\  D {\bf 63}, 103510 (2001)
  [arXiv:astro-ph/0006373].

\bibitem{BaMuVi}
E.~Babichev, V.~Mukhanov and A.~Vikman,
  JHEP {\bf 0802}, 101 (2008)
  [arXiv:0708.0561 [hep-th]].

\bibitem{DMSS}
D.~Maity, S.~SenGupta and S.~Sur,
  Phys.\ Lett.\  B {\bf 643}, 348 (2006)
  [arXiv:hep-th/0604195];
  arXiv:hep-th/0609171;\\
A.~Dey, D.~Maity and S.~SenGupta,
  Phys.\ Rev.\  D {\bf 75}, 107901 (2007)
  [arXiv:hep-th/0611262].

\bibitem{KoMaPe}
L.~Kofman, J.~Martin and M.~Peloso,
  Phys.\ Rev.\  D {\bf 70}, 085015 (2004)
  [arXiv:hep-ph/0401189].

\bibitem{DvGaPo}
 G.~R.~Dvali, G.~Gabadadze and M.~Porrati,
  Phys.\ Lett.\  B {\bf 485}, 208 (2000)
  [arXiv:hep-th/0005016].

\bibitem{CDLPQW}
  M.~S.~Carena, A.~Delgado, J.~D.~Lykken, S.~Pokorski, M.~Quiros 
  and C.~E.~M.~Wagner,
  Nucl.\ Phys.\  B {\bf 609}, 499 (2001)
  [arXiv:hep-ph/0102172].

\bibitem{DvGaKoNi}
  G.~R.~Dvali, G.~Gabadadze, M.~Kolanovic and F.~Nitti,
  Phys.\ Rev.\  D {\bf 64}, 084004 (2001)
  [arXiv:hep-ph/0102216].

\bibitem{BCLPS}
  R.~Bao, M.~S.~Carena, J.~Lykken, M.~Park and J.~Santiago,
  Phys.\ Rev.\  D {\bf 73}, 064026 (2006)
  [arXiv:hep-th/0511266].

\bibitem{CaTaWa}
  M.~S.~Carena, T.~M.~P.~Tait and C.~E.~M.~Wagner,
  Acta Phys.\ Polon.\  B {\bf 33}, 2355 (2002)
  [arXiv:hep-ph/0207056].

\bibitem{Ky}
  B.~s.~Kyae,
  arXiv:hep-th/0207272.

\bibitem{DaHeRi}
  H.~Davoudiasl, J.~L.~Hewett and T.~G.~Rizzo,
  Phys.\ Rev.\  D {\bf 68}, 045002 (2003)
  [arXiv:hep-ph/0212279].

\bibitem{AgPeSa}
  F.~del Aguila, M.~Perez-Victoria and J.~Santiago,
  JHEP {\bf 0302}, 051 (2003)
  [arXiv:hep-th/0302023].

\bibitem{AgPeSa2}
  F.~del Aguila, M.~Perez-Victoria and J.~Santiago,
  Acta Phys.\ Polon.\  B {\bf 34}, 5511 (2003)
  [arXiv:hep-ph/0310353].

\bibitem{MuNiPiRu}
  A.~M\"uck, L.~Nilse, A.~Pilaftsis and R.~R\"uckl,
  Phys.\ Rev.\  D {\bf 71}, 066004 (2005)
  [arXiv:hep-ph/0411258].

\bibitem{CsHuMe}
  C.~Csaki, J.~Hubisz and P.~Meade,
  arXiv:hep-ph/0510275.
\end{thebibliography}
\end{document}